\documentclass[11pt,english]{article}

\usepackage[margin=1in]{geometry}
\usepackage[hang,flushmargin]{footmisc}
\usepackage{setspace}
\usepackage{titlesec}
\usepackage{titlecaps}

\usepackage{amsmath}
\usepackage{amssymb}
\usepackage{amsthm}
\usepackage{graphicx}
\usepackage{booktabs}
\usepackage{natbib}
\usepackage[usenames,dvipsnames]{xcolor}
\usepackage{hyperref}
\hypersetup{
    colorlinks=true,
    linkcolor=MidnightBlue,
    citecolor=MidnightBlue,
    urlcolor=MidnightBlue
}
\usepackage{csquotes}
\usepackage{comment}
\usepackage{authblk}
\theoremstyle{plain}
\theoremstyle{definition}
\newtheorem{definition}{Definition}
\usepackage[title]{appendix}
\usepackage{mathtools}

\title{Rethinking Symmetry Principles and Partial Observables With Reference Frames}
\date{ }

\author[1,2]{Nicola Bamonti}
\author[3]{Henrique Gomes}

\affil[1]{Department of Philosophy, Scuola Normale Superiore, Piazza dei Cavalieri, 7, Pisa, 56126, Italy}
\affil[2]{Department of Philosophy, University of Geneva, 5 rue de Candolle, 1211 Geneva 4, Switzerland}
\affil[3]{Oriel College, University of Oxford, OX14EW, United Kingdom}
\begin{document}

\maketitle

\begin{abstract}
Dynamically uncoupled fields are a modelling device used in the literature on reference frames, classical and quantum. We argue that the device is not innocuous. First, uncoupled fields violate Earman's symmetry principle \textbf{SP1}---that every dynamical symmetry is a spacetime symmetry---and the violation, unlike the familiar Newtonian one, cannot be repaired by excising surplus spacetime structure: it arises from a deficit of dynamical structure, not an excess of spatiotemporal structure. {\color{black}This qualifies implicit assumptions of Earman's \textbf{SP1}, while leaving its diagnostic role intact.} Second, {\color{black}uncoupled fields} expose a lacuna in the partial/complete observables programme: only dynamically coupled fields can serve as \textit{bona fide} partial observables, so partial observability is derivative on complete observability. We then reinterpret the standard formalism for partial observables: \textit{qua} measurable, a partial observable is a relational quantity expressed with respect to an uncoupled frame, and it remains gauge-variant. This reinterpretation vindicates Rovelli's claim that partial observables are individually measurable, against Thiemann and Adlam, and separates two properties easily conflated in the generally covariant formulations of theories: invariance under diffeomorphisms and invariance along the gauge flow.
\end{abstract}

\section{Introduction}\label{sec:intro}

Reference frames have moved to the centre of recent foundational work on general relativity (GR) and quantum gravity. In this literature a reference frame is not a coordinate chart but a physical system---typically a set of matter degrees of freedom---used to localise other fields. It is also standard practice to approximate: one keeps the frame as a localising device while switching off its stress-energy, its dynamics, or both. Such \lq{}dynamically uncoupled\rq{} frames appear in some current classical and quantum treatments \citep[e.g.][]{Rovelli1991,QuantumHole,BamontiCQG}.\footnote{The criterion of dynamical coupling, made formal in \S\ref{sec:coupling}, is this: two fields are dynamically coupled iff the dynamics constrains them jointly---iff which configurations of one are admissible depends on the configuration of the other.} 

This paper argues that the approximation has substantive, and largely unnoticed, consequences in two foundational arenas.

The first concerns symmetry principles. Earman's \citeyearpar{Earman1992-jn} principle \textbf{SP1} requires that every dynamical symmetry of a theory be a spacetime symmetry. {\color{black}While retaining its original role as a diagnostic of surplus spacetime structure, we show that models containing dynamically uncoupled fields violate \textbf{SP1}, and that the violation differs in kind from the textbook Newtonian one. There, surplus spacetime structure shrinks the spacetime symmetry group, and excising the surplus restores the principle. Here, the dynamical symmetry group is enlarged by the absence of coupling, and there is no surplus structure to excise. We therefore make explicit the coupling assumption that \textbf{SP1} tacitly relies on.}

The second concerns the partial and complete observables programme of \cite{Rovelli_2002} and \cite{Dittrich2006,Dittrich2007}. The same analysis shows that the correlation of two dynamically uncoupled fields is a relational but gauge-\textit{variant} quantity. Hence uncoupled frames cannot serve as \textit{bona fide} partial observables: partial observables must be dynamically coupled to their partners, and partial observability is derivative on complete observability. Far from undermining the programme, this refinement strengthens it. Reinterpreting the coordinates in Rovelli's definition of a partial observable as uncoupled physical frames reconciles the gauge-variance of partial observables with their measurability, vindicates their \textit{individual} measurability against \cite{Thiemann:2007} and \cite{adlamobservability}, and separates two properties that diff-invariant theories invite us to conflate: invariance under active diffeomorphisms and invariance along the gauge flow generated by the constraints.

Section \ref{sec:obs} reviews relational observables, introduces the taxonomy of reference frames we rely on, and presents the partial/complete observables framework. {\color{black}Section \ref{sec:symm} defines dynamical coupling, introduces generalised dynamical symmetries, and examines their relation to \textbf{SP1}.} Section \ref{sec:partial} draws the consequences for partial observability. Section \ref{sec:conclusion} concludes.

\section{Relational Observables and Reference Frames}\label{sec:obs}

\subsection{Observables in general relativity}\label{sec:obs-gr}

A physical theory attaches semantics to syntax. Among the mathematical objects of a theory's formalism, some are singled out as representing features of the physical world; these are the theory's \textit{observables}, associated with---but not reduced to---the results of measurements.

The distinction between a variable and an observable becomes imperative in theories with gauge freedom. In Hamiltonian theories this freedom appears as first-class constraints on the variables. \cite{Dirac2012-oj} (see also \citealp{Henneaux1994-ka}) defined observables---now called \textit{Dirac observables}---as quantities that weakly commute with all first-class constraints, or equivalently, that take a single value on each gauge orbit. The two characterisations coincide because the action of the constraints via the Poisson bracket generates infinitesimal gauge transformations, so commutation implies gauge-invariance.

In GR, given a three-dimensional foliation of spacetime, the first-class constraints generate spatial diffeomorphisms along the leaves and deformations normal to them (`refoliations'). For spacetimes satisfying the Einstein field equations (EFEs), these Hamiltonian symmetries correspond neatly to the four-dimensional diffeomorphisms of spacetime \citep{Lee1990}. In what follows, \lq{}gauge-invariant observable\rq{} and \lq{}Dirac observable\rq{} are therefore used synonymously, in the standard sense of constancy along the gauge flow generated by the first-class constraints.

Explicit Dirac observables are hard to find, particularly if they are required to be \textit{spatiotemporally local}. Geometrical objects are anchored to manifold points, the gauge group of GR shuffles those points, and so quantities represented locally in terms of points are not gauge-invariant. This has motivated the widespread view that non-trivial Dirac observables in GR are \textit{non-local}.\footnote{To be precise, \cite{Torre1993} showed that in spatially closed solutions to the vacuum EFEs, no non-trivial local functional of the metric and its conjugate momenta commutes with all first-class constraints. \lq{}Local\rq{} in Torre's sense is a technical term, covering both point-like spacetime functions and integral functionals $F[h_{ij},\pi^{ij}]=\int_{\Sigma^3}{\sigma(x,h,\pi,\partial_x h,\partial_x\pi, \dots)}$ over finite spacelike sub-regions, with $\sigma$ a three-form depending on canonical variables and their derivatives up to some finite order \textit{at} $x \in \Sigma^3$. For asymptotically flat \textit{open} sub-regions with $\partial \Sigma^3 \neq 0$ there \textit{are} non-trivial observables, the ADM observables, and they count as local in this sense (ibid., p.~2). But the ADM observables are insufficient to characterise the state of the gravitational field completely: hence the \textit{problem of local observables} in vacuum GR.\label{fntorre}} Yet non-local observables form too small a class: they do not capture the local gravitational degrees of freedom, and so cannot provide a complete set of observables that uniquely characterises the physical situation \citep{Komar}.

In the presence of matter, however, or by exploiting curvature invariants (e.g.\ the Komar--Kretschmann scalars: \citealp{Kretschmann1917,Kretschmann1918,Komar}), one can construct non-trivial Dirac observables as \textit{relational observables}, whose spatiotemporal localisation is fixed not by bare manifold points but by points as designated by a local \textit{reference frame}.\footnote{The term \lq{}relational\rq{} has a fairly clear meaning in everyday language, but it does not have a single, precise definition. We shall not explore the various possible definitions of this term, but instead adopt a narrow one: a quantity is relational if it correlates fields with different field-theoretic functional roles---a \lq{}dressed object\rq{} on one side, a \lq{}frame\rq{} on the other---expressed covariantly via Eq. \eqref{relationalobs} below.} Such observables are \lq{}relationally local\rq{}.\footnote{They are typically \textit{non-local} in Torre's sense (fn.~\ref{fntorre}). See \cite{Wallace2024,BamontiCQG}, where the non-locality of reference frames, due to their dynamics, is discussed in detail.}

\subsection{Reference frames and their dynamical status}\label{sec:obs-frames}

A \textit{reference frame} is here defined as a set of four scalar degrees of freedom $\Phi:=\{\phi^{(I)}\}|_{I=1,\dots,4}$, associated with a physical (typically, but not necessarily, material) system, that establish a \textit{local} diffeomorphism\footnote{Global reference frames are usually unavailable, for topological obstructions similar to Gribov obstructions \citep{Gribov1978}. Even when the topology allows them, reference frames may generically develop caustics, due to their dynamical nature.\label{fngribov}}
\begin{equation}
    \Phi: U \subset \mathcal{M} \longrightarrow V \subset \mathbb{R}^{4},\qquad
    p \longmapsto \big(\phi^{(1)}(p),\phi^{(2)}(p),\phi^{(3)}(p),\phi^{(4)}(p)\big).\label{mapframe}
\end{equation}
A minimal \textit{viability} requirement on the mapping is local invertibility,
\begin{equation}
    \det\!\left(\frac{\partial \phi^{(I)}}{\partial x^{\mu}}\right)\neq 0
    \quad \text{in some open region } U\subset\mathcal{M},\label{jacobian}
\end{equation}
so that the local inverse $\Phi^{-1}$ allows any field quantity $T_{[ab\dots]}$ to be re-expressed \textit{relationally} as
\begin{equation}
    T_{[IJ\dots]}(\Phi):=[(\Phi^{-1})^{*}]T_{[ab\dots]},\label{relationalobs}
\end{equation}
where $[\bullet]^{*}$ denotes the pullback by $\Phi^{-1}$ and $T_{[ab\dots]}$ is a covariant tensor, understood locally as a section of an appropriate bundle over $\mathcal{M}$.\footnote{We use abstract index notation \citep{Penrose1984} to stress that $T_{[ab\dots]}$ is a geometrical object, independent of any choice of coordinate representation.} Quantities of the form \eqref{relationalobs} are \textit{relational observables}.\footnote{More precisely, whether this relational quantity is also an \textit{observable}, in the sense of being gauge-invariant, depends on the dynamical coupling between $T$ and the reference frame. Establishing this is the burden of \S\ref{sec:partial}.}

Frames differ in how much of their physics the model retains. Following the classification of \cite{Bamonti2023}, refined in \cite{BamontiCQG}, we distinguish four classes in GR, ordered by increasing dynamical involvement:

\begin{enumerate}
    \item \textit{Idealised Reference Frames} (\textbf{IRFs}): both the frame's stress-energy tensor (in the EFEs) and its equations of motion (EOMs) are neglected. An \textbf{IRF} is an \lq{}instantiated\rq{} coordinate chart: it has a physical referent, but one represented in a maximally approximated way. An \textbf{IRF} is not a coordinate chart, however, and the difference matters for what follows. A chart $\varphi:\mathcal{U}\subseteq{\mathcal{M}}\to \mathbb{R}^4$ is fixed as a map on the underlying \lq{}bare\rq{} open set $\mathcal{U}$: it does not covary with active diffeomorphisms. The fields instantiating an \textbf{IRF}, by contrast, are pushed forward and pulled back like any other field---they are \textit{equivariant} under reshufflings of manifold points. Also, coordinates are  \lq{}definitionally\rq{} uncoupled from the fields of the theory; an \textbf{IRF} acquires its uncoupled status through an approximation procedure.\footnote{For details of what this approximation procedure entails in canonical terms, see \cite[\S4.2]{BamontiCQG}.}

    \item \textit{Auxiliary Reference Frames} (\textbf{ARFs}): the frame has well-defined EOMs, but they are governed by an \textit{auxiliary} metric $h_{ab}$, itself dynamically uncoupled from the metric $g_{ab}$ of interest. For example, four Klein-Gordon scalars $\phi^{(I)}$ may propagate on $h_{ab}$ without back-reacting on $g_{ab}$: their evolution is then fixed by initial data and $h_{ab}$ alone, with $g_{ab}$ playing no role.\footnote{The auxiliary metric may be treated either as fixed background structure---an absolute object in the sense of \cite{Anderson1967-en}---or as a dynamical field in its own right. Either treatment is compatible with the \textbf{ARF} classification: what matters is only that the dynamics of $\phi^{(I)}$ is governed by $h_{ab}$ rather than $g_{ab}$. We remain neutral on this choice throughout.}

    \item \textit{Dynamical Reference Frames} (\textbf{DRFs}): only the frame's stress-energy is neglected. The frame is acted upon by the metric but does not act on it: its back-reaction is dropped, but its EOMs, governed by $g_{ab}$, are retained. A realistic example is the set of \textit{GPS coordinates} of \cite{RovelliGPS}. 

    \item \textit{Real Reference Frames} (\textbf{RRFs}): both the frame's dynamics and its stress-energy are retained. Examples are pressureless dust \citep{Brown1995, Giesel2010} and massless scalar fields \citep{Rovelli1994,Domagaa2010}.
\end{enumerate}

This paper needs only a coarser classification. \textbf{IRFs} and \textbf{ARFs} interact in no way with $g_{ab}$; we group them as \textit{Uncoupled Reference Frames} (\textbf{URFs}). \textbf{DRFs} and \textbf{RRFs} are grouped as \textit{Coupled Reference Frames} (\textbf{CRFs}). The absence of back-reaction does not undermine the coupling of a \textbf{DRF}: since its $\phi^{(I)}$ satisfy EOMs governed by $g_{ab}$, which $\phi^{(I)}$ are admissible varies with the metric; in particular, a diffeomorphism acting on $\phi^{(I)}$ alone, with $g_{ab}$ held fixed, generically takes a solution to a non-solution. The criterion is met without the frame acting back on the geometry.

\subsection{Partial and complete observables}\label{sec:obs-partial}

One influential way of formulating relational observables is the framework of \cite{Rovelli1991,Rovelli_2002}, later rigorously formalised by \cite{Dittrich2006,Dittrich2007}. Arguing that gauge-invariant quantities are not the only quantities of physical interest in GR, Rovelli distinguishes two notions of observability. A \textit{partial observable} is a physically instantiated but gauge-variant quantity for which a measurement outcome can be established; it describes the \lq{}phenomenology\rq{}. A \textit{complete observable} is, in the present usage, a relationally constructed, gauge-invariant quantity whose value (or probability distribution, in the quantum case) the theory uniquely predicts. The idea is to relate gauge-variant partial observables to one another in a gauge-invariant manner: a complete observable is a correlation among partial observables.\footnote{Constructing a complete observable requires inverting the gauge flow to locate where the clocks read the prescribed value; in general this can be done only implicitly, as a formal power series, possibly multivalued and only locally convergent. Only in favourable cases, such as clocks evolving linearly in the gauge-flow parameter \citep{Brown1995}, does the series sum to a closed form. The pull-back construction of eq. \eqref{relationalobs} used in our work avoids the series altogether: starting from a solution, with $\Phi$ invertible by our viability assumptions, the frame is used directly as a chart, and the complete observable is read off. The object so obtained is the one \cite[\S8]{Dittrich2006} constructs canonically for gravity coupled to four scalar fields; we reach it covariantly because the dynamics has already been solved.}

\lq{}Complete observable\rq{} and \lq{}Dirac observable\rq{} should not be used interchangeably. Whether a quantity is a Dirac observable is a question about gauge-invariance. Whether it is a complete observable is a question about how it was arrived at, by \textit{relating} quantities that are not gauge-invariant on their own. The global \textit{spacetime} volume $V_g:=\int_\mathcal{M}\sqrt{-g}\,\mathrm{d}^4x$ is constant along the gauge flow, hence a Dirac observable, but no frame enters it and there is nothing it correlates. The four-volume of $\mathcal{M}$ delimited by two readings of a global frame would be constructed as a complete observable; $V_g$ above is not.

The framework suits the view that the physical content of GR lies in relations between dynamical quantities---represented by partial observables---and that we observe relational evolution between fields, not evolution with respect to unobservable background structure such as the bare points of $\mathcal{M}$. Reference frames make the relational idea concrete: the frame supplies one partial observable, and whatever the frame localises supplies the other. Which frames can legitimately play this role is the question of \S\ref{sec:partial}; answering it requires the apparatus of dynamical coupling, to which we now turn.

\section{Symmetry Principles and Uncoupled Fields}\label{sec:symm}

\subsection{Earman's SP principles}\label{sec:SP}

Since this paper deals primarily with general-relativistic theories, we articulate Earman's symmetry principles in terms of internal and external parameters (cf.\ \citealp{Earman1992-jn,Gomes2023b}). \textit{External parameters} are the independent variables; in our case study, the points of the smooth manifold $\mathcal{M}$. \textit{Internal parameters} are the value spaces, described by functions $F$ of the independent variables.

All our models share the form $\langle\mathcal{M},F_n\rangle$, with $n$ a generic index listing the functions. Models in which the $F_n$ are constrained only algebraically---not as (implicit) solutions of differential equations---are \textit{Kinematically Possible Models} (KPMs).\footnote{This generally implies that the $F_n$ can be represented locally as sections of appropriate vector bundles. The space of KPMs may carry further structure (topological, symplectic, etc.).} The subset of KPMs satisfying the dynamical EOMs of the theory are its \textit{Dynamically Possible Models} (DPMs) \citep{James_Read2023-mk}. Whether a given condition counts as algebraic or differential---and hence whether the KPMs or the DPMs absorb it---depends on the choice of variables.\footnote{With the gauge potential $A_a$ as basic variable, the homogeneous Maxwell equation $\mathrm{d}F=0$ holds identically, since $F=\mathrm{d}A$, and constrains no KPM. With the field strength $F_{ab}$ as basic variable, the same $\mathrm{d}F=0$ is a differential condition, and joins the dynamical laws selecting the DPMs. The KPM/DPM partition is thus representation-relative.}

Characterising the symmetries of a spatiotemporal theory requires distinguishing dynamical from background structure. We take the background structure $B$ of a class of models to be whatever the formulation holds fixed across the class \citep[cf.][]{Pooley2015-POOBID,GomesSophLim}: (i) the kinematical structure---here the smooth manifold $\mathcal{M}$, more precisely the smooth structure induced by its maximal atlas (for this \lq{}chart-nominalist\rq{} construal see \citealp{Lang1999}; for the alternative, on which $\mathcal{M}$ carries an intrinsic smooth structure, see \citealp{kobayashi1963foundations}; \citealp{Wallace2019} discusses both)---together with (ii) any fields held fixed across the DPMs, neither varied in the action nor subject to EOMs: the Minkowski metric of a special-relativistic theory, or the auxiliary metric of an \textbf{ARF} when so treated (\S\ref{sec:obs-frames}).\footnote{Talk of background structure within GR can mislead, since GR is often understood as the background-independent theory \textit{par excellence} (see \citealp{Giulini2007,James_Read2023-mk,BamontiBI}). We do not identify $\mathcal{M}$ as a background \textit{spacetime}; nor is there common agreement on how to define spacetime itself.}

With this apparatus:
\begin{description}
    \item \textbf{Spacetime Symmetries} are the transformations that preserve the background structure: $G_S \equiv \mathrm{Aut}(B)$, the automorphisms of the kinematical structure that also preserve the fixed fields. In GR, where $B$ contains no fixed fields, $G_S = Diff(\mathcal{M})$, the group of (active) smooth diffeomorphisms: smooth reshufflings of points. In a special-relativistic theory, $G_S = Poin(\mathcal{M},\eta)\subset Diff(\mathcal{M})$.

    \item \textbf{Dynamical Symmetries} are the transformations that act on the DPMs and preserve solutionhood: they take solutions to solutions and non-solutions to non-solutions. We assume they form an infinite-dimensional group $G_D$. For the EFEs, $G_D \equiv Diff(\mathcal{M})$.
\end{description}

\cite{Earman1992-jn} then proposes two principles:
\begin{description}
    \item[\textbf{(SP1)}] Any dynamical symmetry is a spacetime symmetry;
    \item[\textbf{(SP2)}] Any spacetime symmetry is a dynamical symmetry.
\end{description}
Jointly, the principles require $G_D = \mathrm{Aut}(B)$---in GR, that the dynamical symmetries be exactly those induced by automorphisms of $\mathcal{M}$.\footnote{The equality $G_D=\mathrm{Aut}(B)$ is what \citet{GomesSophLim} calls \textit{background-relative sophistication}; its failure predicts where symmetry must be handled explicitly---gauge-fixing, constraints, dressings---rather than left implicit.} In GR both principles are expected to hold. We will show that \textbf{SP1} contains a tacit assumption: that all fields appearing in the DPMs are dynamically coupled to one another. Models employing \textbf{URFs} violate precisely this assumption.

\subsection{Dynamical coupling}\label{sec:coupling}
Consider two generic fields $\Theta(p)$ and $\Psi(p)$ on $\mathcal{M}$, instances of the functions $F_n$, so that models read $\langle \mathcal{M}, \Theta,\Psi\rangle$. They may be sets of scalar fields, a metric and a scalar, a tensor field and a vector field. Their nature is immaterial, provided the diffeomorphisms act on them canonically---so that pullbacks like $d^*\Theta$ are well-defined, as the coupling definitions below require. This is what it takes for the fields to be sections of \emph{natural} bundles: bundles carrying a canonical lift of $Diff(\mathcal{M})$.\footnote{See \citet{kolar1993natural}. Tensor and spinor fields are natural in this sense; gauge potentials are only \emph{gauge}-natural, and lie outside the present scope.} Nor need both be dynamical: the pair may equally represent a background structure $(\mathcal{M},\Theta)$ and a dynamical field $\Psi$---as in a special-relativistic Klein-Gordon theory, with $\Theta = \eta_{ab}$ the fixed Minkowski metric---which lets the framework cover theories with fixed spacetime backgrounds.

Since kinematical constraints are algebraic and bind each field separately, the space of KPMs factorises as a product, $\mathcal{K}=\mathcal{K}_\Theta\times\mathcal{K}_\Psi$.\footnote{The factorisation is an assumption, not a triviality: an algebraic constraint can bind fields jointly, as a unit-norm condition $g_{ab}u^au^b=1$ binds the pair $(g,u)$. For the fields considered here the kinematical conditions are field-by-field, and $\mathcal{K}$ factorises.} The dynamics selects the DPMs $\mathcal{D}\subseteq\mathcal{K}_\Theta\times\mathcal{K}_\Psi$, and dynamical coupling concerns how it does so: whether the dynamics constrains the two fields jointly, so that which configurations of $\Psi$ are admissible depends on which $\Theta$ they accompany.

\begin{definition}[\textbf{Dynamical coupling}]\label{def:coupling}
The pair $(\Theta,\Psi)$ is \emph{dynamically uncoupled} iff the space of DPMs is closed under recombination: whenever $\langle\mathcal{M},\Theta,\Psi\rangle$ and $\langle\mathcal{M},\Theta',\Psi'\rangle$ are DPMs, so is $\langle\mathcal{M},\Theta,\Psi'\rangle$. The pair is \emph{dynamically coupled} iff it is not dynamically uncoupled.
\end{definition}

Equivalently: the pair is uncoupled iff $\mathcal{D}$ factorises as the product $\mathcal{D}_\Theta\times\mathcal{D}_\Psi$ of its projections, so that solutionhood imposes a separate condition on each field and no joint condition on the pair.\footnote{If $\mathcal{D}$ is closed under recombination, take $(\Theta,\Psi)\in\mathcal{D}_\Theta\times\mathcal{D}_\Psi$: by definition of the projections there are $\Psi_0$, $\Theta_0$ such that $\langle\mathcal{M},\Theta,\Psi_0\rangle$ and $\langle\mathcal{M},\Theta_0,\Psi\rangle$ are DPMs, and recombining them gives $\langle\mathcal{M},\Theta,\Psi\rangle\in\mathcal{D}$. The converse is immediate.} The primes carry no transformation: $\Theta'$ and $\Psi'$ are simply the constituents of another DPM; no map relating them to $\Theta$ and $\Psi$ is presupposed. Spelled out as an existential: the pair is \emph{coupled} iff there are DPMs $\langle\mathcal{M},\Theta,\Psi\rangle$ and $\langle\mathcal{M},\Theta',\Psi'\rangle$ whose recombination $\langle\mathcal{M},\Theta,\Psi'\rangle$ is not a DPM---as when a $\phi'$ solving $\Box_{g'}\phi'=0$ fails to solve $\Box_{g}\phi'=0$ (second remark below).

Three remarks. First, the definition invokes no symmetry: it is meaningful for theories with trivial symmetry groups. Coupling in this sense is modal rather than transformational---a matter of which configurations can dynamically coexist. Two scalar fields on a fixed spacetime with interaction potential $\lambda\,\Theta^2\Psi^2$ are dynamically coupled for $\lambda\neq 0$ and uncoupled for $\lambda=0$, whatever the symmetries of the spacetime.

Being modal, the definition is silent on structures held fixed across all models: a fixed field trivially factorises out of $\mathcal{D}$, and the sense in which matter is \lq{}coupled\rq{} to a fixed background metric is functional---the metric appears in the matter's equations---not modal. The modal notion recovers the verdict once the background is allowed to vary across models. This is a further reason to locate fixed fields in the background $B$ (\S\ref{sec:SP}) and to assess coupling among the dynamical fields. In other words, the definition applies when there are at least two \textit{dynamical} fields; a fixed background metric is not such a field. 

Second, coupling is not determination. The definition does not require that, for given $\Theta$, a unique $\Psi$ be admissible; nor does it require back-reaction. Consider a metric $g_{ab}$ satisfying the vacuum EFEs and a Klein-Gordon scalar $\phi$ satisfying $\Box_g\phi=0$, its stress-energy neglected (a \textbf{DRF} constituent, \S\ref{sec:obs-frames}). For fixed $g_{ab}$, every solution of $\Box_g\phi=0$ is admissible, and one might conclude that the pair is uncoupled, since $\langle\mathcal{M},g_{ab},\phi'\rangle$ is a DPM for many $\phi'\neq\phi$. It is not: which $\phi$ are admissible varies with the metric, so recombining solutions across different metrics generically fails---a $\phi'$ solving $\Box_{g'}\phi'=0$ does not solve $\Box_{g}\phi'=0$. Back-reaction is sufficient for coupling, not necessary.

Third, dynamical coupling is established in practice by inspecting the action or the EOMs: two fields are \emph{directly} coupled when they share interaction terms or appear jointly in the same dynamical equations. The inspection is the practical diagnostic, not the definition: the modal notion also registers \emph{mediated} coupling, transmitted along chains of direct coupling. In graph terms, direct couplings are the edges; the sectors below are the connected components. This will matter in \S\ref{sec:partial}. For models with more than two fields, the definition applies to any bipartition of the fields into sectors; a model may then decompose into sectors that are coupled within and uncoupled across. Call a model \emph{totally coupled} when no bipartition of its fields is uncoupled. A single field $\Phi$ \emph{decouples} from the remaining fields $\{\Psi_1,\dots,\Psi_n\}$ iff $\mathcal{D}=\mathcal{D}_{\Psi_1\cdots\Psi_n}\times\mathcal{D}_{\Phi}$. The full product $\mathcal{D}=\mathcal{D}_{\Phi}\times\prod_i\mathcal{D}_{\Psi_i}$ obtains iff every field decouples, and between the full product---every bipartition uncoupled---and total coupling lie the sector decompositions. Sharing a dynamical partner suffices to fuse sectors: two non-interacting scalars propagating on a common \emph{dynamical} metric form a single sector, since each is coupled to the metric---the $\lambda=0$ verdict above is specific to the fixed background, and that the scalars recombine freely over a \emph{common} metric does not make them a sector of their own. (For an \textbf{ARF}, the relevant sector comprises the frame scalars together with their auxiliary metric, when the latter is treated as dynamical: a private partner restores independent sectors.)

Definition~\ref{def:coupling} mentions no transformations. In the generally covariant formulation of theories it nonetheless makes contact with the symmetries, and the argument of \S\ref{sec:violation} requires the sharpened, symmetry-relative form of that contact.\footnote{General covariance is a property of a formulation, and is cheaply had \citep{Kretschmann1917}; no weight rests on a substantive reading, since the argument concerns a fixed formulation---given field content and background $B$---and asks whether its dynamical symmetries outrun $\mathrm{Aut}(B)$. Why the premise is nonetheless generic: if the EOMs are \emph{natural} differential operators in all their arguments \citep{kolar1993natural}, equivariance $E[d^*\Theta,d^*\Psi]=d^*E[\Theta,\Psi]$ sends solutions to solutions. The condition is not necessary---closure of $\mathcal{D}$ is an on-shell property, naturality an off-shell one. Freezing some arguments of a natural operator (the $\eta$ of a special-relativistic theory, an \textbf{ARF}'s $h_{ab}$) cuts the solutionhood-preserving group to the stabiliser of the frozen arguments: exactly the $\mathrm{Aut}(B)$ of \S\ref{sec:SP}.} 
Since $\Theta$ and $\Psi$ are sections of natural bundles, $Diff(\mathcal{M})$ acts on each factor of $\mathcal{K}_\Theta\times\mathcal{K}_\Psi$ separately, and---in the absence of fixed structures (e.g. Minkowski metric $\eta=\text{diag}(-1,1,1,1)$)---the generally covariant form of the EOMs establishes that \emph{diagonal} actions preserve solutionhood: $\langle\mathcal{M},\Theta,\Psi\rangle$ is a DPM iff $\langle\mathcal{M},e^*\Theta,e^*\Psi\rangle$ is, for all $e\in Diff(\mathcal{M})$.
Recombining diagonal copies of a DPM yields models of the form $\langle\mathcal{M},d^*\Theta,f^*\Psi\rangle$; if the pair is uncoupled, Definition~\ref{def:coupling} guarantees that every such model is a DPM.\footnote{For uncoupled pairs each projection $\mathcal{D}_\Theta$, $\mathcal{D}_\Psi$ respects separately $Diff(\mathcal{M})$-invariance---apply the diagonal action and project---so $(d^*\Theta,f^*\Psi)\in\mathcal{D}_\Theta\times\mathcal{D}_\Psi=\mathcal{D}$.} Whether the converse behaviour attends coupling---whether every \emph{non}-diagonal action breaks solutionhood---is a further question, and the case in which it does deserves its own definition:

\begin{definition}[\textbf{Diff-coupling}]\label{def:diffcoupling}
Let $\langle\mathcal{M},\Theta,\Psi\rangle$ be a DPM. The pair $(\Theta,\Psi)$ is \emph{Diff-uncoupled} iff $\langle\mathcal{M},d^*\Theta,f^*\Psi\rangle$ is a DPM for all $d,f \in Diff(\mathcal{M})$. The pair is \emph{Diff-coupled} iff $\langle\mathcal{M},d^*\Theta,f^*\Psi\rangle$ is a DPM only when there is a single $e \in Diff(\mathcal{M})$ with $d^*\Theta=e^*\Theta$ and $f^*\Psi=e^*\Psi$.
\end{definition}

The clause \lq{}a single $e$\rq{} is required because stabilisers are non-trivial for special configurations: if $d^*\Theta=\Theta$ (say, $d$ is an isometry of a metric $\Theta$), then $\langle\mathcal{M},d^*\Theta,\Psi\rangle = \langle\mathcal{M},\Theta,\Psi\rangle$ is trivially a DPM, and no failure of coupling should be read off from that. A definition that simply demanded $\langle\mathcal{M},d^*\Theta,\Psi\rangle$ fail to be a DPM for all $d$ would be falsified by the identity.\footnote{For generic configurations the stabilisers are trivial \citep{fischer1970theorem}, and Diff-coupling reduces to the intuitive gloss: neither field can be independently diffeomorphed, the other held fixed, without violating solutionhood.}

The two definitions relate as follows. Dynamical uncoupling implies Diff-uncoupling, as shown above. Diff-coupling implies dynamical coupling: for $d$ outside the stabilisers, $\langle\mathcal{M},d^*\Theta,\Psi\rangle$ is not a DPM, yet $\langle\mathcal{M},d^*\Theta,d^*\Psi\rangle$ and $\langle\mathcal{M},\Theta,\Psi\rangle$ are---a failed recombination. Neither implication reverses. In particular, the two clauses of Definition~\ref{def:diffcoupling} are contraries, not contradictories, and dynamically coupled pairs can fall in between, when the dynamics of one field is insensitive to part of the other's structure.\footnote{An example: a source-free Maxwell field treated as test matter (stress-energy neglected) on a vacuum metric admitting a proper conformal isometry $d$, $d^*g_{ab}=\Omega^2 g_{ab}$. By the conformal invariance of the four-dimensional Maxwell equations, $(d,\mathrm{Id})$ preserves solutionhood although $d$ is no isometry, so the pair is not Diff-coupled; it remains dynamically coupled, since recombination across conformally unrelated metrics fails. {\color{black}This is why the analysis of \textbf{SP1} below is formulated in terms of Diff-coupling}: the weaker proviso of total dynamical coupling would not exclude such residual non-diagonal symmetries.} 

In the presence of fixed background fields, e.g. two non-interacting scalars on a fixed $\eta$, Definition~\ref{def:diffcoupling} is to be read relative to the background of \S\ref{sec:SP}—replace $Diff(\mathcal{M})$ by $\mathrm{Aut}(B)$ in both clauses. So relativised, the $\lambda=0$ pair is $\mathrm{Aut}(B)$-uncoupled and the $\lambda\neq0$ pair $\mathrm{Aut}(B)$-coupled, as they should be. In case of a \textit{single} scalar in Minkowski, the pair $(\eta,\phi)$ is $\mathrm{Aut}(B)$-coupled.

For the reference frames of this paper the two notions align. The frames' scalars have metric-governed dynamics with no conformal (or similar) degeneracy, so \textbf{DRF}s and \textbf{RRF}s are Diff-coupled to $g_{ab}$ at generic configurations, while \textbf{IRF}s and \textbf{ARF}s are dynamically---hence Diff---uncoupled from it. The coarse \textbf{URF}/\textbf{CRF} classification of \S\ref{sec:obs-frames} is thus grounded in Definition~\ref{def:coupling}; where the argument needs the symmetry-relative form, we invoke Definition~\ref{def:diffcoupling}.

The validity of \textbf{SP1}, we contend, tacitly assumes that the DPMs are \emph{totally Diff-coupled}: every bipartition of their fields into sectors is Diff-coupled. Coupling is necessary for \textbf{SP1}, though not sufficient---the Newtonian counterexample reviewed below violates \textbf{SP1} for independent reasons.

\subsection{A novel violation of SP1}\label{sec:violation}

\citet[p.~45]{Earman1992-jn} distinguishes dynamical ($P$) and background ($A$) elements of a model, keeping the manifold separate as a standalone background structure, and defines a dynamical symmetry as follows:
\begin{quote}
Consider a model $M=\langle \mathcal{M},A_1,A_2, \cdots ,P_1,P_2, \cdots\rangle$ and let $\Phi$ be a diffeomorphism that maps $\mathcal{M}$ onto $\mathcal{M}$. Define $M_\Phi=\langle \mathcal{M},A_1,A_2, \cdots ,\Phi^*P_1,\Phi^*P_2, \cdots\rangle$. Now $\Phi$ will be said to be a dynamical symmetry just in case for any $M \in \mathbb{M}_T$, it is also the case that $M_\Phi \in \mathbb{M}_T$ [\textit{here $\mathbb{M}_T$ represents the set of all DPMs}].
\end{quote}
In our notation:
\begin{description}
    \item[\textbf{(DS)}] $d\in G_D\subseteq Diff(\mathcal{M})$ is a \textit{dynamical symmetry} iff: $\langle\mathcal{M},\Theta,\Psi\rangle$ is a DPM iff $\langle\mathcal{M},d^*\Theta,d^*\Psi\rangle$ is.
\end{description}
Likewise, the spacetime symmetries of \S\ref{sec:SP}, recast in terms of models:
\begin{description}
    \item[\textbf{(SS)}] $d \in G_S\subseteq Diff(\mathcal{M})$ acts as $\langle\mathcal{M},\Theta,\Psi\rangle \rightarrow \langle\mathcal{M},d^*\Theta,d^*\Psi\rangle$, for all KPMs.
\end{description}

\textbf{(DS)} builds the \lq{}diagonal action\rq{} into the very format of a dynamical symmetry: a \textit{single} $d$ acts uniformly on all dynamical fields. Models such as $\langle\mathcal{M},d^*\Theta,\Psi\rangle$ or $\langle\mathcal{M},\Theta,d^*\Psi\rangle$ are thereby excluded from being symmetry-related to $\langle\mathcal{M},\Theta,\Psi\rangle$, whatever the dynamics. This format is harmless so long as the model is totally Diff-coupled, since then, by Definition~\ref{def:diffcoupling}, only diagonal actions (up to stabilisers) preserve solutionhood anyway. But it prejudges the uncoupled case. We therefore generalise:
\begin{description}
    \item[\textbf{(GDS)}] $(d,f) \in G_D\subseteq Diff(\mathcal{M})\times Diff(\mathcal{M})$ is a \textit{generalised dynamical symmetry} iff: $\langle\mathcal{M},\Theta,\Psi\rangle$ is a DPM iff $\langle\mathcal{M},d^*\Theta,f^*\Psi\rangle$ is.
\end{description}
\textbf{(GDS)} introduces no transformations beyond those describable as elements of $Diff(\mathcal{M})$ acting \textit{non-diagonally} on the fields. Its point is conceptual bookkeeping. \textbf{(DS)} makes every dynamical symmetry a single diffeomorphism acting on all fields at once; and where the background is the bare manifold, so that $G_S=\mathrm{Aut}(B)=Diff(\mathcal{M})$, any such diffeomorphism is a spacetime symmetry. {\color{black}Under Earman's format, then, $G_D \subseteq G_S$ holds by construction and \textbf{SP1} is satisfied. The further question whether the dynamical symmetries outrun the spacetime symmetries cannot even be posed.} \textbf{(GDS)} makes it posable; \textbf{(DS)} is recovered as the diagonal subgroup.\footnote{Earman's \lq{}just in case\rq{} would already admit a more encompassing reading if interpreted as an \lq{}if\rq{} rather than an \lq{}iff\rq{}: there may be transformations that preserve solutionhood yet cannot be written as a single $d$ acting diagonally, and the \lq{}iff\rq{} reading excludes them from the class of dynamical symmetries by definition rather than by argument.}

The violation of \textbf{SP1} is now immediate. Consider a model $\langle\mathcal{M}, g_{ab}, \phi\rangle$, where $g_{ab}$ satisfies the EFEs and $\phi$ is a \textbf{URF}---for definiteness, an \textbf{IRF}: a set of four scalars with neither stress-energy in the EFEs nor EOMs of its own. The pair $(g_{ab},\phi)$ is dynamically uncoupled in the sense of Definition~\ref{def:coupling}: solutionhood concerns only $g_{ab}$'s equations, so the space of DPMs factorises as $Sol(\text{EFE})\times\mathcal{K}_\phi$; and since the EFEs are generally covariant, $\langle\mathcal{M}, d^*g_{ab}, f^*\phi\rangle$ is a DPM for all $d,f \in Diff(\mathcal{M})$ (\S\ref{sec:coupling}). In particular, taking $f=\mathrm{Id}$, the pair $(d,\mathrm{Id})$ is a \textbf{(GDS)}: it relates the DPMs $\langle\mathcal{M}, g_{ab}, \phi\rangle$ and $\langle\mathcal{M}, d^*g_{ab}, \phi\rangle$ while preserving solutionhood. But $(d,\mathrm{Id})$ is not a spacetime symmetry: a spacetime symmetry, being a reshuffling of the underlying points of $\mathcal{M}$, acts through a \textit{single} diffeomorphism on \textit{all} fields defined on $\mathcal{M}$, as \textbf{(SS)} says. So here is a dynamical symmetry that is not a spacetime symmetry: an explicit violation of \textbf{SP1}. When uncoupled fields are in play, the dynamical symmetries are no longer \lq{}\lq{}just a reflection of the automorphisms of $\mathcal{M}$\rq{}\rq{} \citep{Gomes2023}.

The example does not require both fields to be dynamical: the \textbf{IRF} $\phi$ has no dynamics at all, and \textbf{(GDS)} applies as soon as at least one field has equations whose solutionhood is to be preserved. Nor is the example special to two fields: any model that partitions into mutually uncoupled sectors, each with generally covariant equations, admits independent diffeomorphisms sector by sector. Two Klein-Gordon fields propagating on distinct auxiliary metrics (two \textbf{ARFs}), for instance, may be diffeomorphed independently provided each sector---scalar together with its auxiliary metric---is dragged as a whole.

{\color{black} This example leaves Earman's \textbf{(SP1)}, formulated with \textbf{(DS)} intact. It identifies a  a novel way to highlight the limits of \textbf{SP1}'s applicability, showing the consequences of using uncoupled fields. Our argument thus clarifies the scope of the correspondence between dynamical and spacetime symmetries. Also, the sense of the violation is precise: it gives no reason to replace Earman's diagnostic.}

\subsubsection*{Why the violation is novel}

Violations of \textbf{SP1} are not unprecedented, and Earman has a standard diagnosis of them:
\begin{quote}
    The theory that fails \textbf{SP1} is thus using more space-time structure than is needed to support the laws, and slicing away this superfluous structure serves to restore \textbf{SP1}. \citep{Earman1992-jn}
\end{quote}
The textbook case is Newtonian mechanics: Newtonian spacetime has a standard of absolute rest, a structure not preserved by the Galilean boosts that are dynamical symmetries of the EOMs. \textbf{SP1} fails because surplus spacetime structure has \textit{shrunk} $G_S$, and it is restored by reformulating the theory in Galilean spacetime---slicing the surplus away.

Nor is the Newtonian case the only precedent. Violations of the same formal type arise whenever a formulation introduces fixed structure that the dynamical symmetries do not preserve \citep{GomesSophLim}: in linearised gravity, the perturbation $h_{ab}$ propagates on a rigid $\eta_{ab}$, and the gauge shifts $h_{ab}\mapsto h_{ab}+\mathcal{L}_\xi\eta_{ab}$ are dynamical symmetries---read variationally, as transformations preserving the action up to boundary terms---that are no automorphisms of $(\mathcal{M},\eta)$; the potential formulation of electromagnetism, with its shifts $A_a\mapsto A_a+\partial_a\chi$, behaves analogously. These cases still answer to Earman's diagnosis, suitably generalised: the background $B$ has been enriched, $\mathrm{Aut}(B)$ has shrunk below $G_D$, and the repair excises or de-rigidifies the offending structure---de-linearise, or pass to a formalism in which the gauge symmetry acts as an automorphism of the background \citep{GomesSophLim}.

\begin{figure}[htbp]
    \centering
    \includegraphics[width=0.5\textwidth]{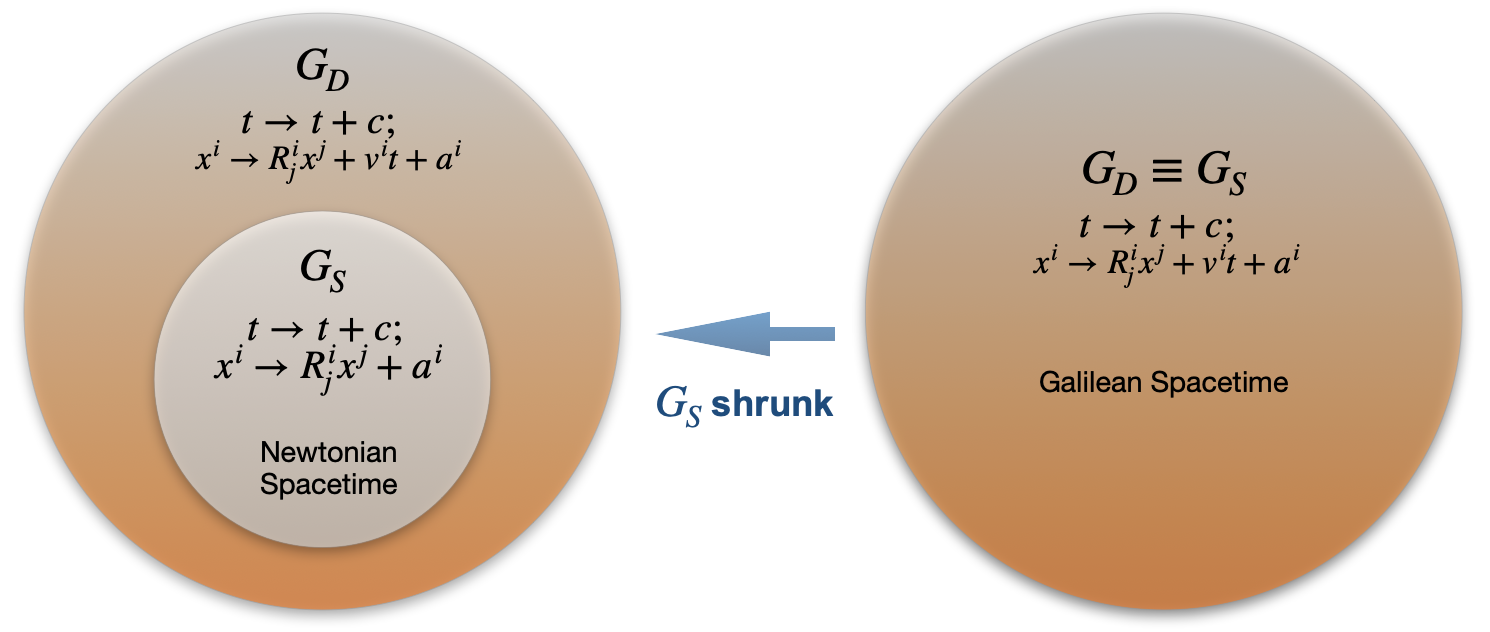}
    \includegraphics[width=0.42\textwidth]{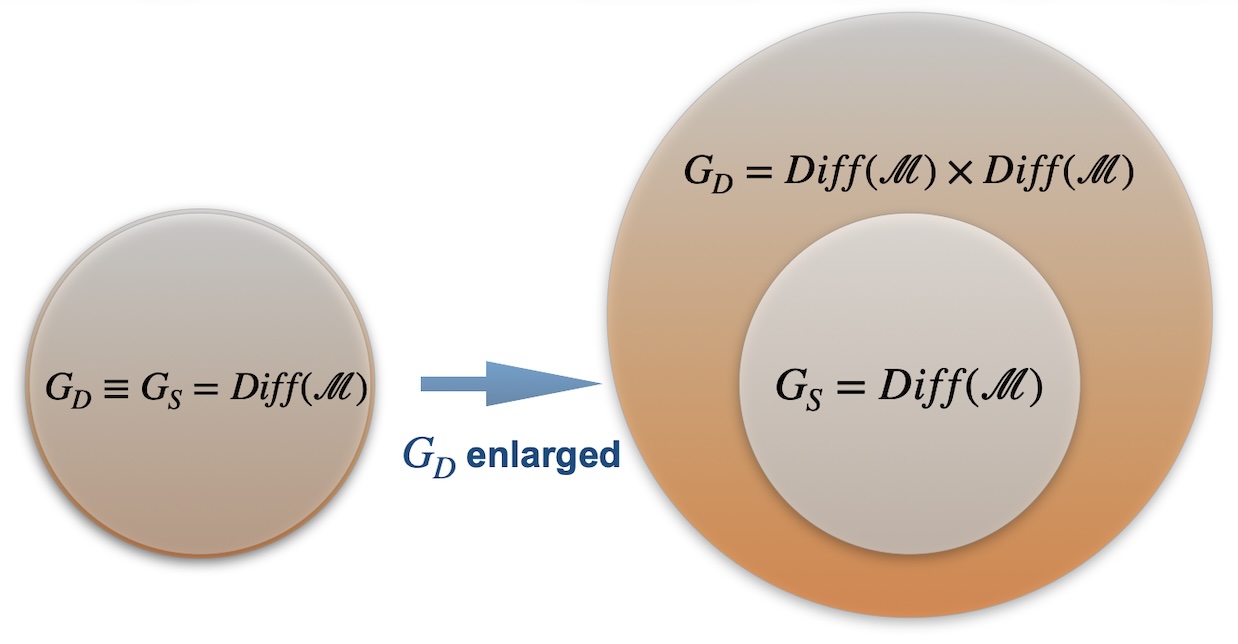}
    \caption{Left: the textbook violation of \textbf{SP1}, in which the spacetime symmetries shrink from the Galilean group to Newton's translations. Right: the violation presented in this section, in which the dynamical symmetry group $G_D$ is extended. Symbols: $t$ the coordinate time; $c$ a constant; $x^i$ the spatial coordinates of a point; $R^i_j$ a $3\times3$ orthogonal rotation matrix; $v^i$ the velocity; $a^i$ a spatial translation vector.}
    \label{figSP1}
\end{figure}

Our violation fits neither diagnosis.\footnote{\textcolor{black}{The violation pertains to the original definition of \textbf{SP1} and its implicit assumption regarding coupling. It would not be a violation if one were to consider that the definition applies only to a coupled field, that is what our work make explicit.}} It arises not by adding background structure ($\mathrm{Aut}(B)$ shrinking) but by removing dynamical structure ($G_D$ enlarging): see Fig.~\ref{figSP1}. The models of \S\ref{sec:violation} carry no fixed fields, so there is nothing to de-rigidify; and with $G_S = \mathrm{Aut}(B) = Diff(\mathcal{M})$ already the largest group available short of arbitrary point permutations or merely continuous transformations, there is nothing to slice away either---weakening the smooth structure to a topological one would not bridge the gap to $G_D$.\footnote{This is also why we ran the argument on an \textbf{IRF}. If the auxiliary metric of an \textbf{ARF} is treated as a fixed field, it belongs to the background (\S\ref{sec:SP}), $\mathrm{Aut}(B)$ shrinks accordingly, and models containing \textbf{ARF}s violate \textbf{SP1} twice over: once in Earman's fashion, through the fixed $h_{ab}$, and once in ours, through the uncoupling from $g_{ab}$. The \textbf{IRF} keeps the violation pure.} The surplus symmetries also differ in character: there, field-shifts that are not diffeomorphism actions at all; here, honest diffeomorphisms, merely acting sector by sector. Restoring \textbf{SP1} therefore requires \textit{adding dynamical structure}---coupling the frame: a change of theory, not of formulation.\footnote{One may insist that if two sectors are totally uncoupled, we could represent them on distinct manifolds $\mathcal{M}_1$, $\mathcal{M}_2$. Then $\mathrm{Aut}(B)=Diff(\mathcal{M}_1)\times Diff(\mathcal{M}_2)$, sector-wise diffeomorphisms are automorphisms of the background, and \textbf{SP1} is restored: the surplus structure sliced away is the co-location of the sectors on a single manifold. Such repair is unavailable for the models at issue. A \textbf{URF} earns its keep by localising the other sector's fields: the relational quantity \eqref{relationalobs} presuppose that frame and target are defined on a common domain $\mathcal{M}$.}

\textbf{SP2}, by contrast, {\color{black}is not violated}. Given $\langle\mathcal{M},g_{ab},\phi\rangle$, every model $\langle\mathcal{M},d^*g_{ab},d^*\phi\rangle$ related to it by a spacetime symmetry is also related by a (generalised) dynamical symmetry, since the diagonal action preserves the solutionhood of generally covariant equations and the \textbf{URF} simply equivaries. Hence $G_S \subset G_D$: the inclusion that fails is the converse one.

\section{What Is a Partial Observable, Anyway?}\label{sec:partial}

The choice of dynamical coupling has repercussions beyond symmetry principles: it bears directly on the partial/complete observables programme, since coupling is what grounds the \textbf{URF}/\textbf{CRF} distinction, and frames are the natural candidates for partial observables. This section establishes two claims. First (\S\ref{sec:bonafide}): only dynamically coupled fields can serve as \textit{bona fide} partial observables, so \textbf{URFs} cannot, and partial observability is derivative on complete observability. Second (\S\ref{sec:relational}--\ref{sec:measurement}): the standard formalism for partial observables should be reinterpreted---a partial observable, \textit{qua} measurable, is a relational quantity expressed with respect to a \textbf{URF}, and remains gauge-variant. The reinterpretation separates diff-invariance from gauge-invariance and underwrites the individual measurability of partial observables against the challenge of Thiemann and Adlam.

Throughout, unless otherwise stated, one partial observable is a material reference frame (a set of four scalar fields), serving as the standard of spatiotemporal localisation for the other, a dynamical field on $\mathcal{M}$ (typically a general-relativistic metric). Their relation is a relational observable in the sense of eq.~\eqref{relationalobs}---not yet, we will argue, a complete one.

\subsection{Bona fide partial observables are dynamically coupled}\label{sec:bonafide}

Partial observables are standardly introduced by association with a measurement outcome (\S\ref{sec:obs-partial}). The literature rarely emphasises that partial observables must be \textit{dynamically coupled} for their relation to form a complete, hence gauge-invariant, observable. Rovelli, analysing the significance of coordinates in GR, takes what we have classified as \textbf{URFs} (he calls them undetermined physical coordinates) to be partial observables:
\begin{quote}
    The coordinates are partial observables in (i) physical coordinates, with an interpretation as positions with respect to objects whose equations of motions are taken into account and (ii) physical coordinates with an interpretation as positions with respect to objects \textit{whose equations of motions are ignored} [our italics]. \citep[p.~63]{Rovelli2004}
\end{quote}
We read \lq{}ignored\rq{} as meaning that the reference objects are dynamically \textit{uncoupled} from the other dynamical objects of the theory, through an approximation procedure (see \citealp{BamontiCQG}).\footnote{We believe the quote contains a typo. In light of what \citeauthor{Rovelli2004} writes on p.~62---\lq{}\lq{}[the physical] coordinates $X^\mu$ are interpreted as the spacetime location of reference objects whose dynamics we \textit{have chosen} to ignore\rq{}\rq{}---clause (ii) should read \lq{}\lq{}positions \textit{of} objects whose equations of motions are ignored\rq{}\rq{}: it is the dynamical equations \textit{of} the reference objects that are ignored.}

The claim about (ii) is false, or so we now argue. \textit{Bona fide} partial observables must be dynamically coupled to each other for their relation to constitute a complete observable; the correlation between dynamically uncoupled fields is, at most, a gauge-variant relational observable. It is not enough to take any two measurable quantities and correlate them: the quantities must \textit{already} be dynamically coupled. This is in line with \citet[p.~9]{adlamobservability}: \lq{}\lq{}partial observables are simply how complete observables look `from the inside,' so to speak.\rq{}\rq{} The procedure is top-down: one obtains two partial observables by \lq{}unpacking\rq{} a complete observable.\footnote{One might worry that this top-down reading is epistemically circular: identifying two fields as \textit{bona fide} partial observables seems to require already knowing they combine into a complete observable, while constructing the complete observable seems to presuppose identifying the right coupled partial observables. The circularity is only apparent. Dynamical coupling is not a property of individual quantities, to be detected one quantity at a time; one establishes it by inspecting the action or the EOMs---whether the fields share interaction terms or appear jointly in the same dynamical equations (\S\ref{sec:coupling}). The top-down order reflects the logical priority of the dynamics over the subsequent choice of observable, not a need to know the complete observable before identifying its parts.\label{fnadlam}}

We now show that \textbf{CRFs} can play the role of \textit{bona fide} partial observables and \textbf{URFs} cannot. Suppose we have a metric field $g_{ab}$ satisfying the EFEs and a set of four massless Klein-Gordon scalars $\{\phi^{(I)}\}$ used as a (local) reference frame; the pair $(g_{ab},\phi^{(I)})$ is a particular case of the $(\Theta,\Psi)$ of \S\ref{sec:coupling}. For \textit{any} such pair, the metric can be parametrised by the values of the scalars as in eq.~\eqref{relationalobs}: we write the relational metric as $g_{IJ}(\Phi):= \big[\Phi^{-1}\big]^* g_{ab}$, the pulled-back components of the abstract tensor $g_{ab}$ in the frame $\{\phi^{(I)}\}$.\footnote{By our viability assumptions, any two value-configurations of the scalar fields are related by a diffeomorphism of $\mathcal{U}\subset \mathcal{M}$, since $\Phi^{-1}\circ \Phi'\in Diff(\mathcal{U})$.} Whether $g_{IJ}(\Phi)$ is a complete observable depends on whether $\{\phi^{(I)}\}$ is a \textbf{URF} or a \textbf{CRF}. The deciding condition is gauge-invariance, so we analyse the two cases in terms of initial values and constancy along gauge orbits, taking the relevant group to be $Diff(\mathcal{U})$, $\mathcal{U}\subset \mathcal{M}$.

\begin{enumerate}
 \item[i)] \textbf{URFs.} By the argument of \S\ref{sec:violation}, any $\phi^{(I)}$ is compatible with \textit{all} the diffeomorphic copies of the metric: both $\langle \mathcal{M}, g_{ab}, \phi^{(I)}\rangle$ and $\langle \mathcal{M}, [d^*g]_{ab}, \phi^{(I)}\rangle$ are DPMs, for all $d \in Diff(\mathcal{U})$ acting non-diagonally. We have no reason to start with $g_{ab}$ rather than any $[d^*g]_{ab}$, even granting a reason to start with one $\phi^{(I)}$ rather than its diffeomorphic copies.\footnote{The same holds with the roles reversed: a given metric is compatible with all the diffeomorphic frames $\langle \mathcal{M}, g_{ab}, d^*\phi^{(I)}\rangle$.} That is, we have no reason to prefer $\big[\Phi^{-1}\big]^* g_{ab}$ over $\big[\Phi^{-1}\big]^*[ d^* g]_{ab}$. So $g_{IJ}(\Phi)$ is not gauge-invariant: an effective action of $Diff(\mathcal{U})$ remains, and we are left with the same redundancy we started with when considering the bare $g_{ab}(p)$. A \textbf{URF} therefore cannot play the role of a \textit{bona fide} partial observable with respect to $g_{ab}$: a field serves as a partial observable only if its correlation with its partner yields a \textit{unique} complete observable, constant along the gauge flow of the first-class constraints.

\item[ii)] \textbf{CRFs.} Here the pair is Diff-coupled (Definition~\ref{def:diffcoupling}): when $\langle\mathcal{M}, g_{ab},\phi^{(I)}\rangle$ is a DPM, $\langle\mathcal{M},[d^*g]_{ab},\phi^{(I)}\rangle$ is not, for $d \in Diff(\mathcal{U})$ acting non-diagonally; only the diagonal action, relating $(g_{ab},\phi^{(I)})$ and $([d^*g]_{ab},d^*\phi^{(I)})$, preserves solutionhood. For given initial conditions, a choice of $\phi^{(I)}$ (rather than any of its diffeomorphic copies) yields a \textit{unique, on-shell} representation $g_{IJ}(\Phi)$: there is no other choice that stays on-shell. The choice of frame completely fixes the gauge, securing gauge-invariance.\footnote{For the debate on whether choosing a frame is different from gauge-fixing, see \cite{Berghofer2024} (in favour of the distinction) and \citet[\S4.5]{BamontiCQG} (against). See also \cite{deAGomes2025}, where choosing a frame is presented as choosing a section of the principal fibre bundle of models of the theory.} Thus $\phi^{(I)}$ (and $g_{ab}$) can play the role of \textit{bona fide} partial observables.
\end{enumerate}

\subsection{Partial observables are relational, yet gauge-variant}\label{sec:relational}

\textbf{URFs} cannot yield complete observables; local \textbf{CRFs} can. Building on this result, we now reinterpret the formalism standardly used to define a partial observable.

\cite{Rovelli_2002} formally defines a partial observable as a measurable, gauge-variant quantity expressed in a coordinate chart $\{x^\mu\} \in \mathbb{R}^4$---for instance, the familiar gauge-variant metric components $g_{\mu\nu}(x^\rho)$.\footnote{In \cite{Dittrich2006,Dittrich2007}, partial observables (for a finite-dimensional mechanical system with a single Hamiltonian constraint) are phase-space functions $T(x;\tau)$, with $x \in \Gamma=\mathbb{R}^N\times \mathbb{R}^N$ a phase-space point and $\tau$ the gauge parameter of the constraint flow. They are individually measurable quantities defined with respect to a gauge parameter, not relations between physical systems: the formal definition is not relational.\label{fndittrich}} But coordinates are mathematical artefacts without physical instantiation. How can partial observables, so defined, be empirically measurable?

Empirical measurement outcomes are relational: they result from interactions between physical systems---naively, the measured object and the instrument \citep{Giovanelli2021}. We accordingly read the requirement that partial observables be measurable as requiring that they be relationally given. The needed reinterpretation must therefore make partial observables relational while preserving their characteristic gauge-variance. \textbf{URFs} make this possible. The relational metric $g_{IJ}(\Phi)$, with $\{\phi^{(I)}\}$ a \textbf{URF}, is relational by construction and, by \S\ref{sec:bonafide}(i), gauge-variant.

It is also invariant under active diffeomorphisms, as a short calculation shows:
\begin{align}
d^*[g_{IJ}(\Phi)]:=\big[(d^*\Phi)^{-1}\big]^* [d^*g]_{ab}&=\big[(\Phi \circ d)^{-1}\big]^* [d^*g]_{ab}\notag
\\&= \big[d^{-1} \circ \Phi^{-1}\big]^* [d^*g]_{ab}\notag
\\&= \big[\Phi^{-1}\big]^* \circ \big[d^{-1}\big]^*[d]^*g_{ab} =: g_{IJ}(\Phi).\label{diffinvproof}
\end{align}
So $g_{IJ}(\Phi)$, with $\{\phi^{(I)}\}$ a \textbf{URF}, is \textit{diff-invariant} yet \textit{gauge-variant}. 
Invariance under active diffeomorphisms---which eq.~\eqref{diffinvproof} delivers for any frame, coupled or not, simply because both relata equivary---is one property. Constancy along the gauge flow generated by the first-class constraints---which, by \S\ref{sec:bonafide}, requires dynamical coupling---is another. Since in GR the gauge group and the diffeomorphism group are standardly identified, this invites the conflation of the two; the relational quantities built from \textbf{URFs} prise them apart.\footnote{We are not claiming, conversely, that all diff-invariant quantities are relationally constructed via the pullback formalism adopted here.} A recent instance of the conflation is \citet[p.~22]{thiemann26}, which states that \lq{}\lq{}relational observables are gauge-invariant by construction\rq{}\rq{}: eq.~\eqref{diffinvproof} together with \S\ref{sec:bonafide}(i) shows that construction delivers diff-invariance for free, but gauge-invariance only given coupling.

Accordingly, the coordinate chart in \cite{Rovelli_2002}'s definition of a partial observable should be interpreted as a \textbf{URF}.\footnote{Analogously, the gauge parameter of the constraint-flow equation in \cite{Dittrich2006,Dittrich2007} (fn.~\ref{fndittrich}) acts as a \textbf{URF}. See also \cite{BamontiThebault2024}.} So interpreted, partial observables are measurable, because relationally given.

We have claimed both that a partial observable is relationally represented with respect to a \textbf{URF} (this section) and that a \textit{bona fide} partial observable must be a \textbf{CRF}, dynamically coupled to its partner (\S\ref{sec:bonafide}). The two claims concern different things. Coupling characterises a field's \textit{dynamical} relation to its partner---a fact about the EOMs, indifferent to representation. Relational representation characterises how the field is \textit{presented} as a measurement outcome---a fact about modelling, indifferent to the field's dynamical role. A field coupled to its partner can itself be expressed relationally with respect to a third, uncoupled field. The toy model of \S\ref{sec:measurement} exhibits exactly this situation.

\subsection{Measuring partial observables individually}\label{sec:measurement}

The reinterpretation just proposed also supports Rovelli's account against a well-known line of attack, on which the partial/complete distinction is misguided because the \textit{only} measurable quantities are complete observables. The position was first championed by \cite{Thiemann:2007}, who rejects the distinction in favour of a single notion of \textit{observable simpliciter}, encapsulating both Dirac's technical definition and the operational definition of a measurable quantity:\footnote{See also \citet[p.~21]{thiemann26}: \lq{}\lq{}A basic postulate in gauge theories is that experiments have only access to gauge-invariant functions on the phase space that are called observables\rq{}\rq{}.}
\begin{quote}
A measurable quantity is always a complete observable, even pointers of a clock are observables and not partial observables. Now complete observables are defined with respect to non-measurable quantities...which we will simply call non-observables (ibid., p.~78).
\end{quote}
\cite{adlamobservability} has recently brought the position back to the fore, in a more conciliatory register:
\begin{quote}
We surely cannot measure complete relational observables by measuring two partial observables separately and then comparing their values; we must somehow measure the complete relational observable all together [\dots] So a natural way to think about Rovelli's proposal is to say that an observer is able to observe partial observables relativized to her body or location, because she is thereby measuring a complete relational variable pertaining to the relation between her and the entity she is measuring (ibid., pp.~8--9).
\end{quote}
Adlam keeps the distinction but attributes measurability strictly to complete observables, never to the partial observables composing them.

Against this \lq{}Thiemann--Adlam approach\rq{}, we defend Rovelli's position that partial observables \textit{can} be individually measured. The defence must be reconciled with \S\ref{sec:bonafide}: partial observables are defined as such precisely by being dynamically coupled parts of a complete observable, so partial observability is a concept \textit{derivative} on complete observability. Our strategy is to introduce a more liberal, \textit{model-based} notion of measurement \citep{Tal2013TheEO,tal:2016,BamontiThebault2024}, naturally suited to relational partial observables. We take a measurement to consist of three stages:\footnote{The three-stage division is not the most refined possible; it is the one that serves our purpose. Actual measurement procedures are complex and vary considerably from case to case.}
\begin{enumerate}
\item \textit{Raw data acquisition}: collecting direct signals from the experimental apparatus (in cosmology, say, fluxes and redshifts);
\item \textit{Modelling the measured quantity}: formally modelling the measured quantity as the result of an interaction between coupled physical systems (instrument/frame and observed system). \textbf{CRFs} model \textit{realistic} measurements;
\item \textit{Uncoupling}: approximating away the dynamical coupling of the frame employed in the measurement,\footnote{Other approximations are of course made; we focus on the one that eliminates the frame's dynamical role.} so that the \textbf{CRF} is replaced by a \textbf{URF}. The resulting model exhibits gauge-variance as a by-product of the approximation. \textit{Only in this liberal sense does one \lq{}measure a relational partial observable\rq{}}.
\end{enumerate}
Realistic measurement procedures inherently involve dynamically coupled fields; measurement \textit{outcomes} are nonetheless often represented by relational, gauge-variant quantities. This liberal notion of measurability partially reconciles Thiemann's and Rovelli's positions: strictly speaking, only gauge-invariant observables are measured (Thiemann), yet partial observables are measurable in the specified sense (Rovelli). The reconciliation has limits. Thiemann's claim is not that, as a matter of contingent practice, only complete observables happen to be measured; it is the stronger claim that measurability is \textit{constitutively} tied to gauge-invariance, so that calling a gauge-variant quantity \lq{}measurable\rq{}---in however liberal a sense---is a category error. A committed Thiemannian may therefore reject the three-stage model at the outset, on the grounds that liberalising the notion of measurement is precisely what is in dispute. Our defence of Rovelli rests on accepting that measurement outcomes can be relational and gauge-variant when expressed with respect to a \textbf{URF}; readers who refuse that premise will not be persuaded. Symmetrically, a committed Rovellian will not be persuaded by restricting measurability to gauge-invariant quantities by fiat.

\begin{figure}
    \centering
    \includegraphics[width=0.43\linewidth]{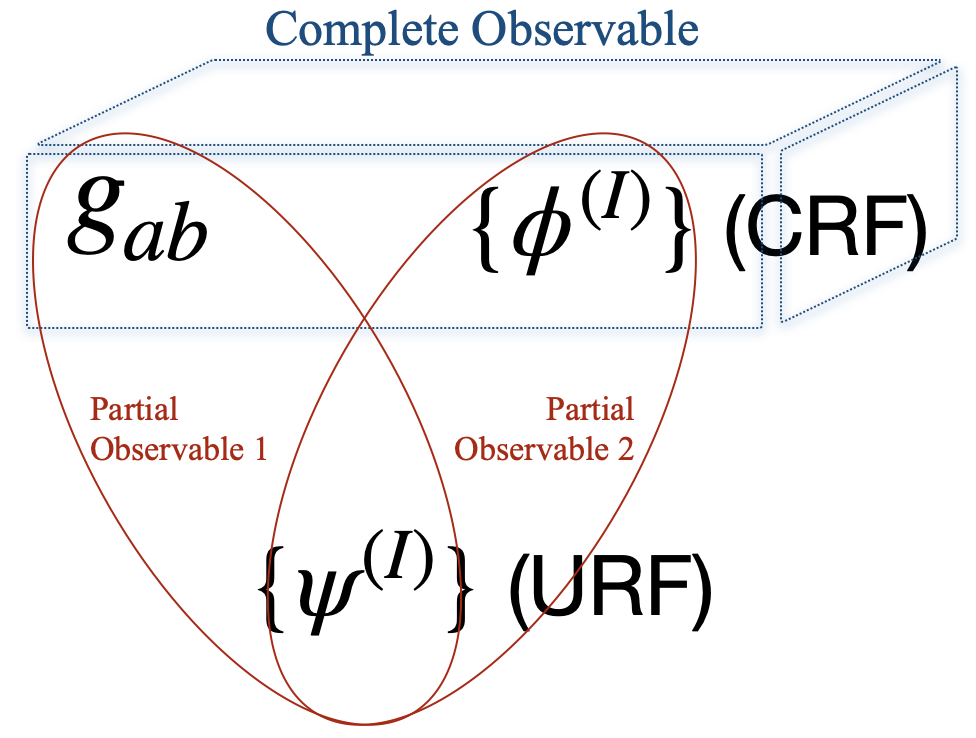}
    \caption{A complete observable constituted by two dynamically coupled, \textit{relational} partial observables, each independently measurable. The box around the complete observable illustrates that the two partial observables are obtained by \lq{}unpacking\rq{} it.}
    \label{toymodel}
\end{figure}

One step remains. Measuring a relational, gauge-variant quantity defined with respect to a \textbf{URF} does not yet amount to measuring a \textit{partial observable}: the quantity might simply never be used to construct a complete observable, for want of a partner to which it is dynamically coupled. We must therefore show that the \textit{individual} measurability of two such quantities is compatible with their \textit{bona fide} status as the coupled parts of a complete observable. A toy model makes the compatibility explicit (Fig.~\ref{toymodel}).

Consider four real, massless, free Klein-Gordon fields $\Phi:=\{\phi^{(I)}\}$ constituting a \textbf{CRF} with respect to a metric $g_{ab}$: each scalar satisfies $\Box_g\phi^{(I)}=0$, so $\Phi$ can play the role of partial observable for the metric. Consider also a physically distinct set of four scalars $\Psi:=\{\psi^{(I)}\}$ constituting a common \textbf{URF} with respect to both $g_{ab}$ and $\{\phi^{(I)}\}$; the two sets of scalars do not interact.\footnote{Since $\{\phi^{(I)}\}$ and $\{\psi^{(I)}\}$ are independent free fields, there is no dynamical functional dependence between them. The usual viability assumptions for $\{\psi^{(I)}\}$ as a well-behaved frame apply (\S\ref{sec:obs-frames}).} We construct the two relational observables $g_{IJ}(\Psi)$ and $\phi^{(I)}(\Psi)$, where $\phi^{(I)}(\Psi):=\phi^{(I)}\circ \Psi^{-1}$ denotes the value of each scalar in $\{\phi^{(I)}\}$ as a function of the four scalars in $\{\psi^{(I)}\}$. Both are \textit{individually measurable} in the three-stage sense above. We then combine the two expressions to describe the correlation between the values of $g_{ab}$ and $\{\phi^{(I)}\}$, by the standard procedure for constructing a complete observable: invert (locally) one partial observable---say $\phi^{(I)}(\Psi)$---to obtain each $\psi^{(I)}$ as a function of $\{\phi^{(I)}\}$, and substitute into $g_{IJ}(\Psi)$.\footnote{For even slightly complicated systems the inversion step may run into significant obstacles; \cite{Dittrich2007} treats various cases.} Given the dynamical coupling between $(g_{ab},\phi^{(I)})$, the composition is the complete observable $g_{IJ}(\Phi)$---which exhibits the \textit{bona fide} status of $g_{IJ}(\Psi)$ and $\phi^{(I)}(\Psi)$ as partial observables.

The toy model also displays the dissolution of \S\ref{sec:relational}'s worry in concrete form. The \textbf{URF} $\{\psi^{(I)}\}$ is the \lq{}instantiated coordinate chart\rq{} with respect to which both relational partial observables $g_{IJ}(\Psi)$ and $\phi^{(I)}(\Psi)$ are expressed when individually measured; the dynamical coupling of the pair $(g_{ab},\phi^{(I)})$ makes them \textit{bona fide} partial observables.\footnote{Dynamical coupling governs whether one field's dynamics bears on another's; it says nothing about how a single field is represented (fn.~\ref{fnadlam}).} Representation and dynamical status come apart; the construction needs both.

This completes our defence and refinement of \cite{Rovelli_2002}'s framework.

\section{Conclusion}\label{sec:conclusion}

The dynamical status of a reference frame---coupled or uncoupled to the fields it localises---is not a technical detail but a modelling choice with substantive consequences for two pillars of the foundations of GR.

{\color{black}For symmetry principles: dynamically uncoupled fields admit dynamical symmetries that act independently, sector by sector, and such symmetries are not spacetime symmetries. This highlight a violation of Earman's \textbf{SP1} principle in a way his own diagnosis does not account for. The analysis preserves Earman's diagnostic and clarify its scope of applicability, highlighting the consequences of using uncoupled fields. The novelty consists in the violation of \textbf{SP1} stemming from a deficit of dynamical structure, not a surplus of spacetime structure, so there is nothing to slice away. This makes clear that the principle holds provided the DPMs are totally Diff-coupled. Coupling itself required care to define: dynamical coupling is modal---failure of the space of DPMs to factorise---and presupposes no symmetry; its symmetry-relative sharpening, Diff-coupling, is the notion \textbf{SP1} and the observables argument invoke.} The violation is not an idle formal possibility, since \textbf{URFs} are in certain cases employed in the classical and quantum reference-frame literature; treating such fields as genuine physical fields costs a principle Earman regards as central to any adequate spatiotemporal theory.

For observables: only \textbf{CRFs} can serve as \textit{bona fide} partial observables, so partial observability is derivative on complete observability---one unpacks a complete observable into its coupled parts. Reinterpreting the coordinates of the standard formalism as \textbf{URFs} renders partial observables relational, hence measurable, while preserving their gauge-variance; it separates diff-invariance from gauge-invariance, two properties that the generally covariant formulation of theories invite us to conflate; and, with a model-based notion of measurement, it vindicates the individual measurability of partial observables against Thiemann and Adlam---within the limits flagged in \S\ref{sec:measurement}.

We leave for future work the ontological status of uncoupled fields in light of the \textbf{SP1} violation, and the extension of the analysis to the quantum reference-frame setting.

\newpage 
\bibliography{biblio.bib}

\end{document}